\documentclass{iaus}
\usepackage{graphicx}

\title[Formation of young B/PS bulges] 
{The formation of young B/PS bulges in edge-on barred galaxies}

\author[H. Wozniak \and L. Michel-Dansac]   
{H. Wozniak$^1$ \and  L. Michel-Dansac$^2$}

\affiliation{$^1$Observatory of Lyon, F-69561 Saint-Genis-Laval cedex, France \break 
email: herve.wozniak@obs.univ-lyon1.fr\\[\affilskip]
$^2$IATE, Observatorio astronomico, X5000BGR Cordoba, Argentina, \break 
email: leo@oac.uncor.edu}

\pubyear{2008}
\volume{245}  
\pagerange{119--126}
\date{?? and in revised form ??}
\jname{Formation and Evolution of Galaxy Bulges}
\editors{M. Bureau, E. Athanassoula, and B. Barbuy, eds.}
\begin{document}

\maketitle

\begin{abstract}
We report about the fact that the stellar population that is born in
the gas inflowing towards the central regions can be vertically
unstable leading to a B/PS feature \emph{remarkably bluer that the
  surrounding bulge} . Using new chemodynamical simulations we show
that this young population does not remain as flat as the gaseous
nuclear disc and buckles out of the plane to form a new boxy bulge. We
show that such a young B/PS bulge can be detected in colour maps.

\end{abstract}

\begin{figure}
\centering
\includegraphics[scale=0.55]{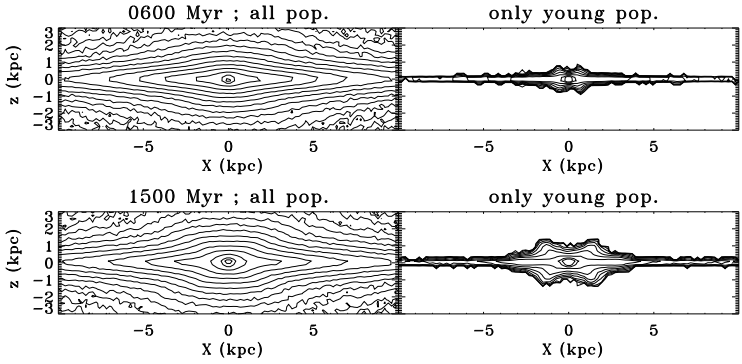}
\caption{Mass distribution (in log units) of the central
  10$\times$5~kpc seen edge-on for T=600 (top panel) and 1500~Myr
  (bottom). Left panel: all stars (old and young populations) are
  plotted. Right panel: only the young population has been plotted.}
\label{fig:mass}
\end{figure}

In the generic case of pure N$-$body simulations, whenever a disc
galaxy forms a bar, a B/PS bulge develops in a few dynamical times.
In the case of chemodynamical simulations, with gas and star
formation/feedback recipes (cf \cite{wmd07,mdw08} for full details),
the B/PS growing process is different and more complicated due to the
presence of a young stellar population that is born in the disc. 
Most of this young population lies in a razor-like central disc during
the first 450~Myr, obviously because of the small vertical scaleheight
of the initial gas distribution. Since a razor-thin disc is highly
unstable the most central part of the disc starts to thicken out the
equatorial plane. In roughly a bar rotation period, the vertical
distribution gets symmetrically peanut shaped over the central 2~kpc,
while the young bar is approximately 8~kpc long (cf. Fig
\ref{fig:mass}, $t=600$ Myr). At this time, the total mass of the
central disc being still low, the thickening process has no clear
detectable effect on the whole mass distribution. Then, the old
population also starts to buckle out leading to a larger B/PS bulge
(cf Fig. \ref{fig:mass}, $t=1500$ Myr). Both populations being fully
mixed, the two components can only be splitted in numerical
simulations.

To observationally detect the young stellar component of a B/PS bulge
we have to rely on stellar population tracers, e.g. B$-$V maps. We
thus obtained mock B$-$V maps calibrating our simulations. The full
process is described by \cite{mdw04}. We assume that the initial
population starts with an age of 10.4 Gyr and has a solar metallicity
at $z=0$.

\begin{figure}
\includegraphics[scale=0.4]{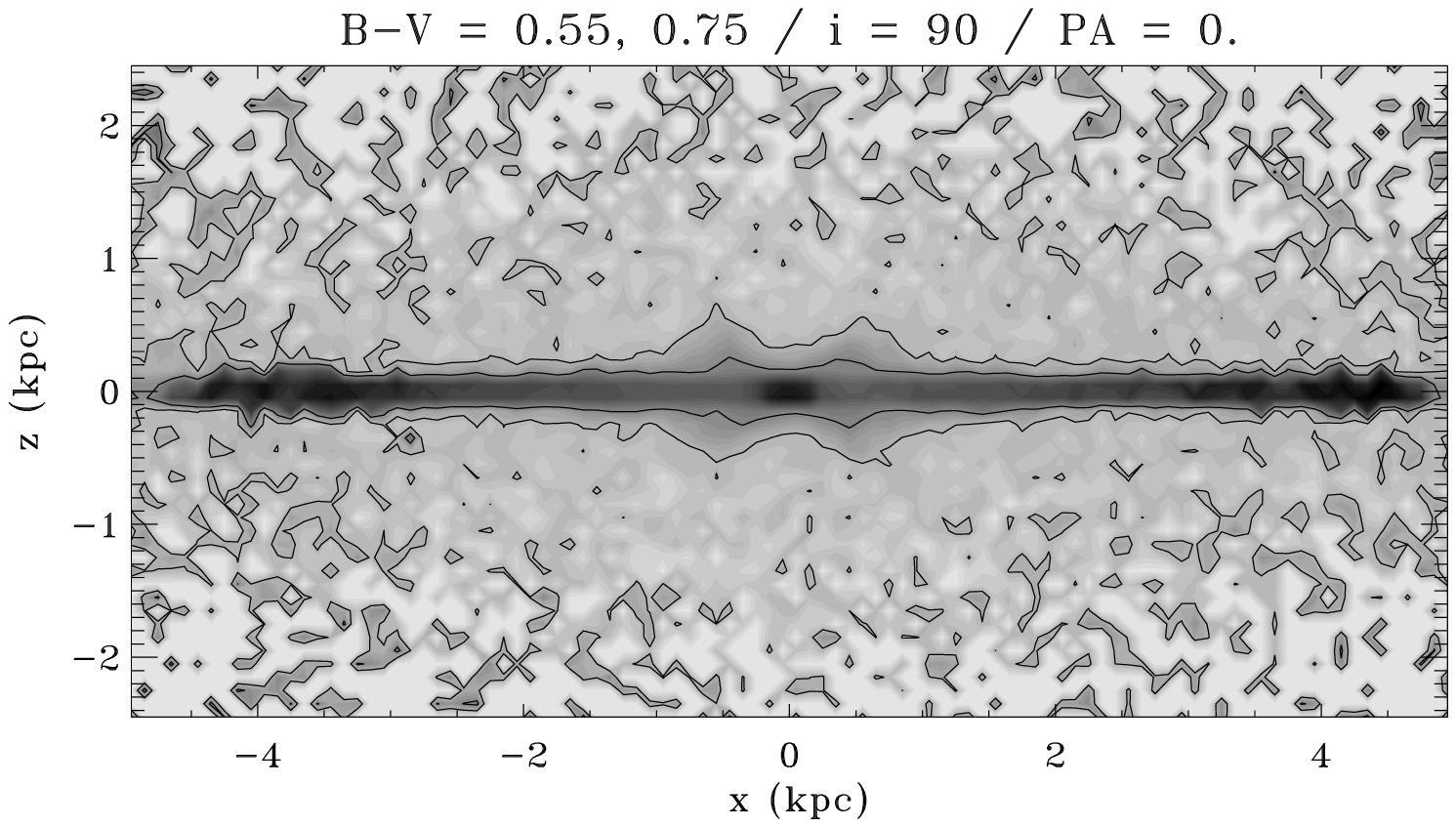}
\includegraphics[scale=0.4]{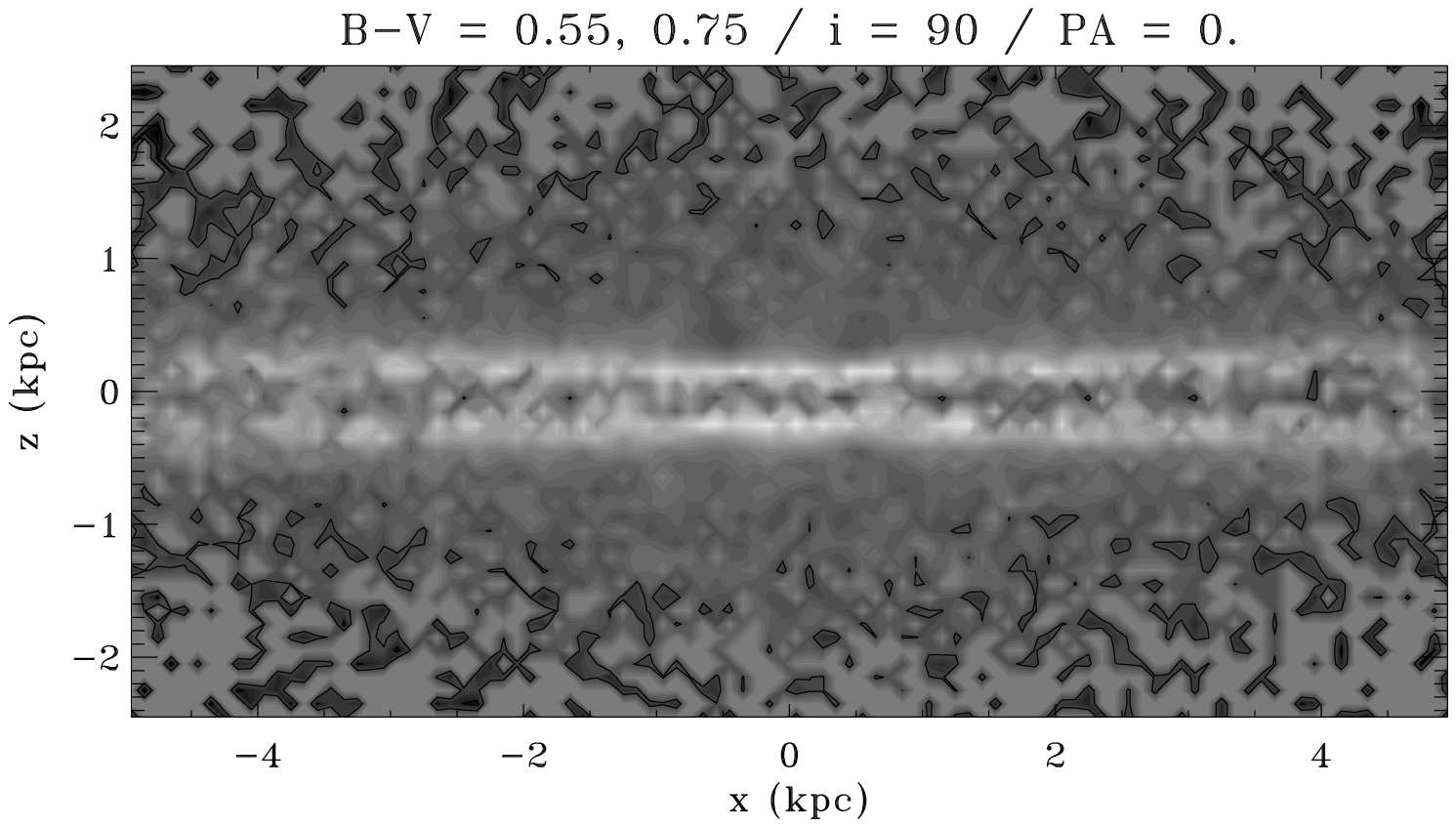}
\includegraphics[scale=0.4]{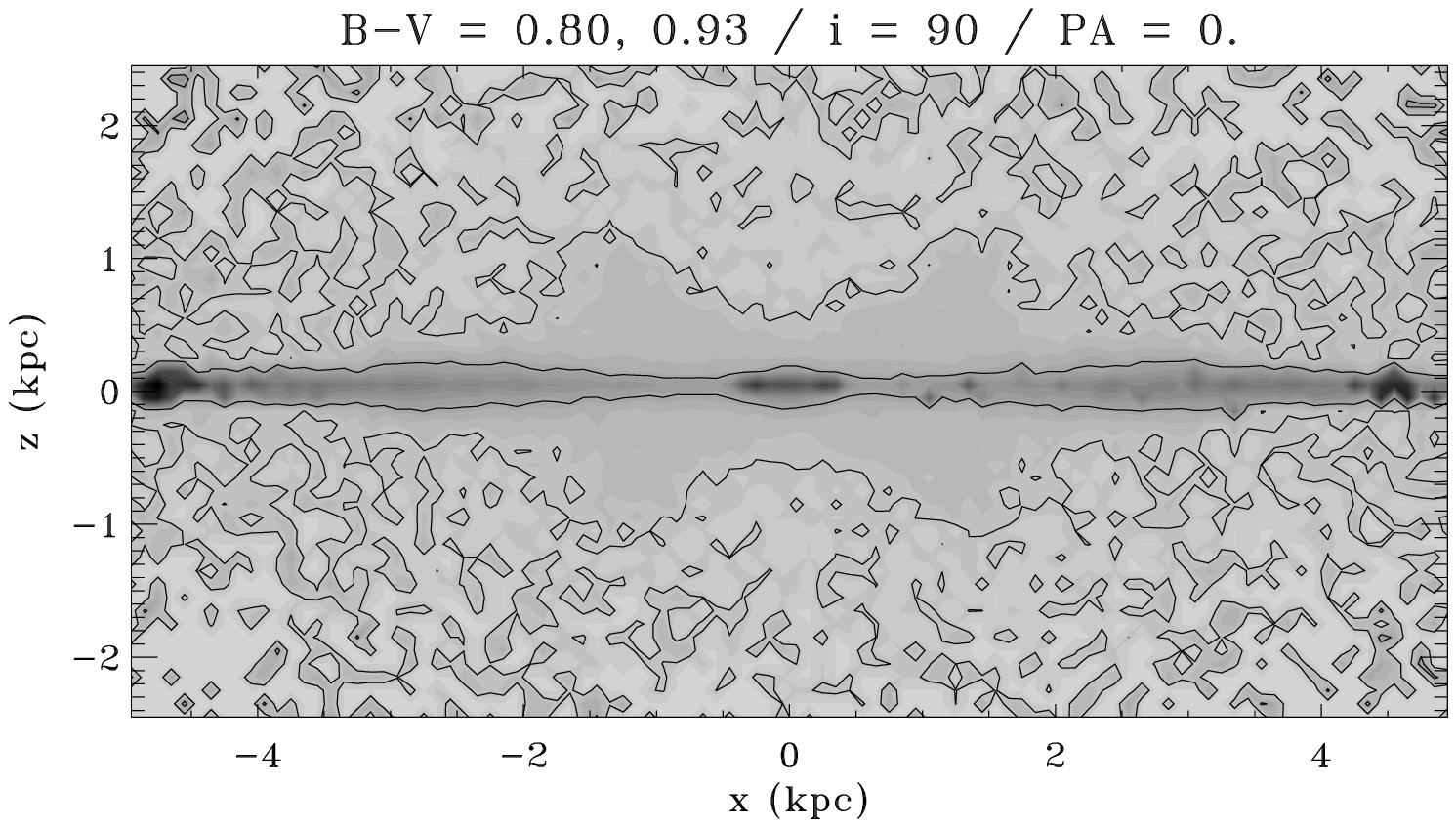}
\includegraphics[scale=0.4]{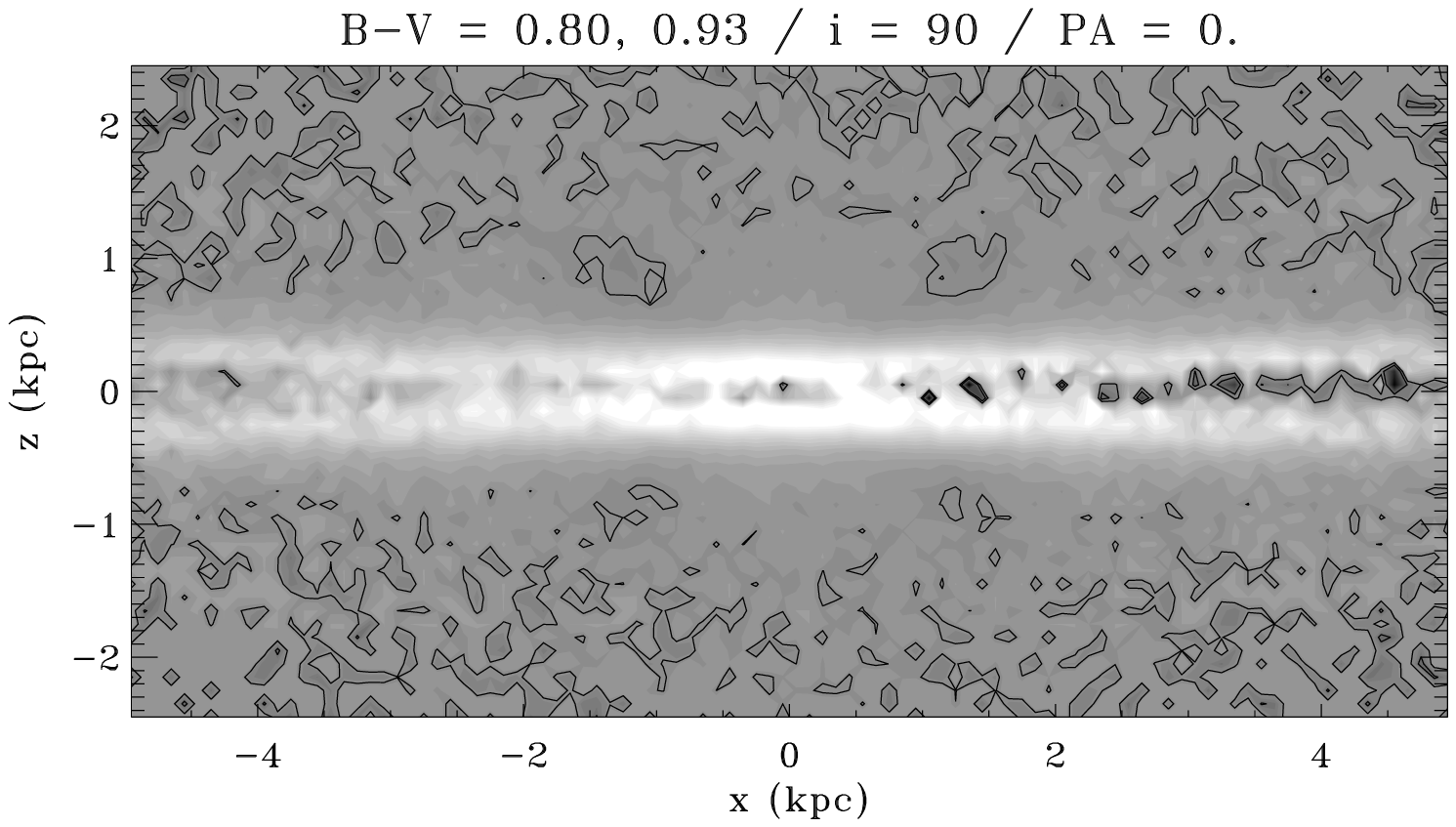}
\caption{B$-$V colour maps without (left) and with (right) a dust
  component for the same fields and times. Bluest region are coded in
  black. The two isocontours have been chosen as to enhance the boxy
  feature and are quoted at top of each frame.}
\label{fig:edgeon}
\end{figure}

Due to the colour contrast between both stellar populations, the young
B/PS structure is clearly visible (Fig.~\ref{fig:edgeon}). The dust
could however hampered such a detection (Fig.~\ref{fig:edgeon}, right
panels). At $t = 600$~Myr, since the new stellar disc is young and
star formation still active, the dust amount should be likely large.
Afterward, the peanut-shape widens out as the young disc evolves. The
thick part of the disc doubles its radial size in less than 1~Gyr. The
vertical scaleheight also increases with time leading to a
well-developed peanut-shaped bulge for $t > 1500$~Myr.

The young B/PS structure, being now much more extended
both in the radial and vertical direction, is likely detectable even
in the presence of dust. Indeed, the height of the B/PS structure is
greater that the dust disc scaleheight. 

Since the early studies of B/PS bulges, there were many evidences that
galaxies hosting such a feature are edge-on barred disc galaxies, and
that the B/PS bulges themselves represent the thickest parts of the
bars. But stellar populations of B/PS bulges (and their colours) have
been rarely studied. Looking for young, blue and small scaleheight
B/PS bulge, likely inside older, redder and larger bulge, deserves
dedicated surveys.

\end{document}